\documentclass[aip,apl,reprint,amsmath,amssymb]{revtex4-2}

\usepackage{graphicx}
\usepackage{bm}
\usepackage[colorlinks=true,linkcolor=blue,citecolor=blue,urlcolor=blue]{hyperref}

\graphicspath{{figures/}}

\newcommand{\Tmat}{\mathbb{T}}
\newcommand{\Einc}{\mathbf{E}^{\mathrm{inc}}}
\newcommand{\Esca}{\mathbf{E}^{\mathrm{sca}}}

\begin{document}

\title{Spectral-Domain Deep Learning of Intrinsic Scattering Operators for Arbitrarily Shaped Compact 3D Particles}

\author{Daize Li}
\affiliation{State Key Laboratory of Intelligent Construction and Healthy Operation and Maintenance of Deep Underground Engineering, Shenzhen University, Shenzhen 518060, China}
\affiliation{Department of Physics and Astronomy ``Galileo Galilei'', University of Padova, Via Marzolo 8, 35131 Padova, Italy}

\author{Jiafu Shen}
\affiliation{State Key Laboratory of Intelligent Construction and Healthy Operation and Maintenance of Deep Underground Engineering, Shenzhen University, Shenzhen 518060, China}

\author{Yifei Liu}
\email{yfliu@szu.edu.cn}
\affiliation{State Key Laboratory of Intelligent Construction and Healthy Operation and Maintenance of Deep Underground Engineering, Shenzhen University, Shenzhen 518060, China}

\author{Bonan Zhang}
\affiliation{College of Physics and Electronic Information, Inner Mongolia Normal University, Hohhot 010022, China}

\author{Heping Xie}
\affiliation{State Key Laboratory of Intelligent Construction and Healthy Operation and Maintenance of Deep Underground Engineering, Shenzhen University, Shenzhen 518060, China}

\date{\today}

\begin{abstract}
Rapid prediction of optical scattering from arbitrarily shaped three-dimensional particles is important for particle optics and photonic characterization, but remains challenging because of the large variability of complex morphologies and the strong angular dependence of their scattering responses.
To address both issues, a dual spectral-domain neural scattering model is introduced in which morphology and scattering are represented in physically ordered bases: particle geometry is compressed into only 256 spherical-harmonic coefficients, and the optical response is encoded by the complex T-matrix in a spherical-vector-wave basis.
The morphology spectrum replaces high-dimensional Euclidean geometry representations, such as voxel grids, point clouds, or meshes, with a compact ordered descriptor, while the T-matrix represents a geometry-determined scattering operator that can be queried for different incidence directions, polarizations, and observation angles.
A spectral-token Transformer trained on 50{,}000 irregular particles at 1064~nm maps the morphology spectrum directly to the T-matrix.
The predicted operators recover modal structure and reproduce full-angle differential scattering maps and incidence-angle scans.
Generalization to out-of-distribution synthetic shapes and natural sand-particle morphologies shows that the dual spectral architecture learns an intrinsic relation from the geometry spectrum to multipolar scattering.
This establishes spectral-domain operator learning as a compact route for reusable, angle- and polarization-resolved optical scattering prediction of complex 3D particles.
\end{abstract}


\maketitle

Electromagnetic scattering by compact irregular particles underpins optical sensing, microscopy, aerosol optics, colloidal photonics, and engineered light--matter interactions. When the particle size is comparable to the wavelength, the response is governed by Mie-regime interference and becomes strongly dependent on morphology, orientation, material, and illumination. Full-wave methods such as finite-element, finite-difference time-domain, boundary-element, and method-of-moments solvers can provide accurate solutions, but repeated simulations over many particle shapes and optical queries remain expensive. This has motivated neural surrogates and neural operators for electromagnetic modeling.\cite{Bayvel:2012electromagnetic,Chen:2018computational,Bondeson:2012computational,Ma:2021deep,Salucci:2022artificial,Liu:2023recent}

A central difficulty for learning-based scattering prediction is how to represent both the shape and the response. Many existing models use high-dimensional Euclidean geometry representations, including voxel grids, meshes, point clouds, or parameterized shapes, and regress scattered fields, far-field patterns, or scalar observables for prescribed optical conditions.\cite{Wu:20153D,Fan:2017point,Wang:2018pixel2mesh,Giannakis:2019machine,Shan:2020study,Ma:2020learning,Yin:2022electric} Such models are useful accelerators, but the learned map can remain tied to the chosen geometry discretization and to the chosen output observable. For irregular three-dimensional particles, a more compact and structured formulation should encode morphology in a common basis while preserving the modal information that determines scattering.

The problem is therefore formulated as a dual spectral-domain map. On the input side, the particle surface is represented by an angular-harmonic morphology spectrum, so complex geometry is compressed into a small ordered coefficient vector rather than a voxel grid, point cloud, or mesh. On the output side, the optical response is represented by the complex T-matrix in a spherical-vector-wave basis, which collects the multipolar coupling structure of the particle at a fixed wavelength. The aim is not simply to replace one solver output with one neural prediction, but to learn the relation between spectral shape content and the finite-dimensional scattering operator.

With these paired representations, model construction becomes a compact spectral translation from ordered shape coefficients to multipolar response coefficients. The morphology side avoids direct learning from geometry-dependent Euclidean discretizations, while the T-matrix side preserves phase, polarization mixing, and angular-momentum coupling that would be lost in a single intensity image or scalar cross section. The resulting surrogate is therefore compact in its description of geometry, expressive in its description of scattering, and naturally aligned with the modal structure of electromagnetic theory.

A Transformer architecture is adopted for this spectral translation because the input is an ordered sequence of spherical-harmonic modes rather than an unordered geometric sample. Self-attention can connect distant spectral orders and learn correlations between global shape modes and finer surface perturbations, in analogy with its use for sequence and vision representation learning.\cite{Vaswani:2017attention,Han:2021transformer}

To evaluate this formulation, 50{,}000 random irregular particles are generated, their boundary-element T-matrices are computed at 1064~nm, and the spectral-token Transformer is trained to learn the map from morphology spectra to T-matrices. The learned map is then validated at three levels: through comparisons of complex T-matrix modal structure, conversion of predicted operators into differential scattering observables, and tests on out-of-distribution synthetic shapes and real sand-particle morphologies to assess whether a transferable relation from geometry spectra to scattering responses has been learned.

\begin{figure*}[t]
    \centering
    \includegraphics[width=0.88\textwidth]{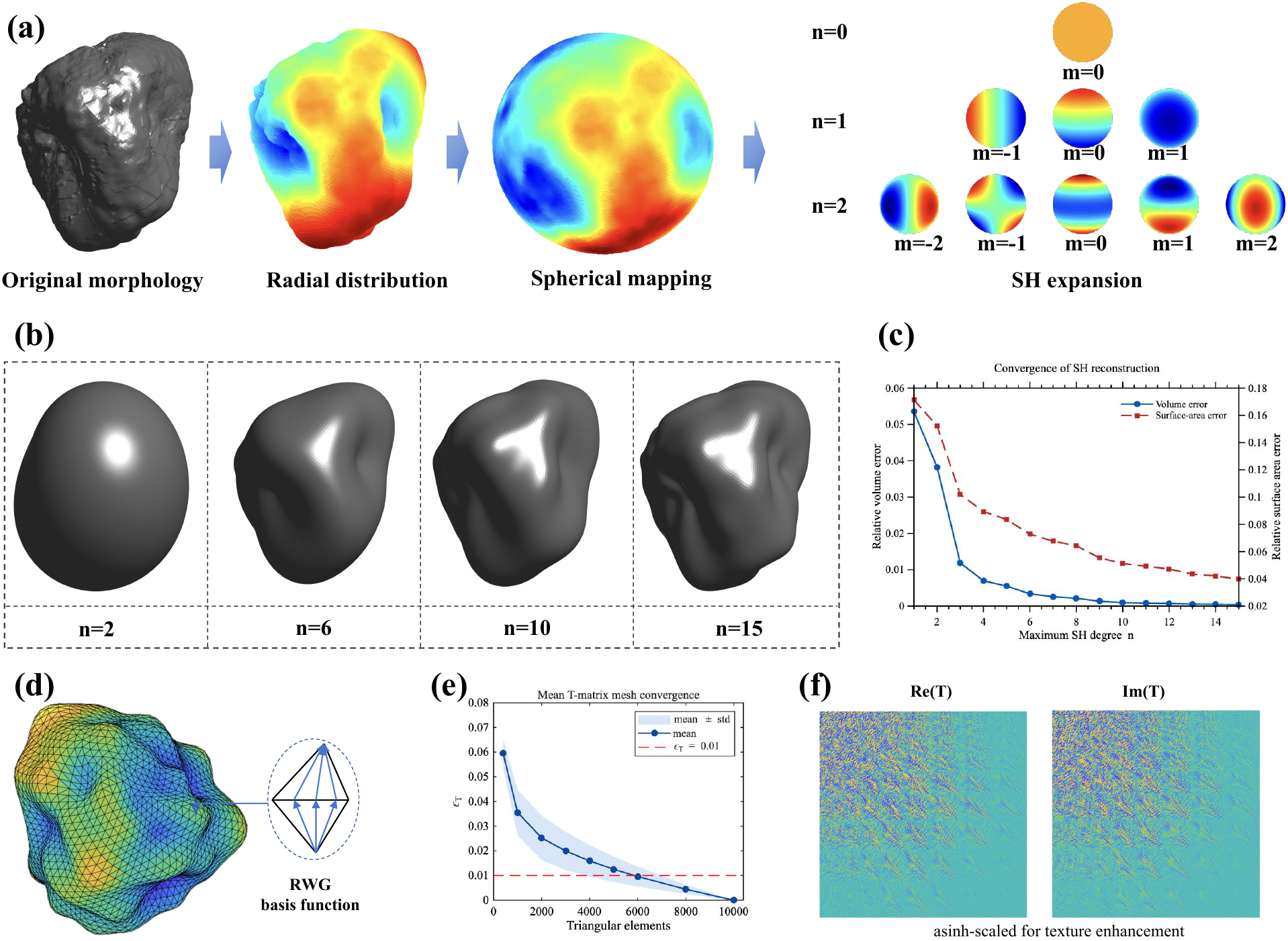}
    \caption{Spectral representation of irregular 3D particle morphology and generation of T-matrix ground truth. (a) A compact star-shaped particle is converted from the original morphology to a radial distribution, mapped onto the unit sphere, and expanded in spherical harmonics. (b) Reconstructions with increasing maximum degree show how low-order modes describe global geometry and high-order modes recover finer surface structure. (c) Relative volume and surface-area errors quantify convergence of the truncated representation. (d) Triangular surface mesh used for boundary-element T-matrix computation, together with the Rao--Wilton--Glisson (RWG) basis function on adjacent triangles. (e) Mean T-matrix error as a function of triangular mesh density, showing convergence toward the adopted tolerance. (f) Representative real and imaginary T-matrix components visualized with asinh scaling to reveal weak modal-coupling texture.}
    \label{fig:sh_representation}
\end{figure*}

The morphology spectrum is defined for compact star-shaped particles whose surfaces are expressed by radial functions on the unit sphere. Each radial function is expanded as
\begin{equation}
r(\theta,\phi)=\sum_{n=0}^{\infty}\sum_{m=-n}^{n}c_n^mY_n^m(\theta,\phi),
\label{eq:sh_shape}
\end{equation}
where \(Y_n^m\) are spherical harmonics and \(c_n^m\) are morphology coefficients determined from the surface point cloud.\cite{Shen:2006large} Here ``spectral'' refers to the angular-harmonic content of the shape, rather than to a wavelength sweep. Low harmonic orders encode global size and elongation, whereas higher orders encode localized surface variations. Truncating at degree 15 yields 256 coefficients and a surface-area reconstruction error below \(10^{-3}\) for the particles used here, providing a strongly compressed description of complex 3D morphology. This compression is a key feature of the framework: instead of learning from high-dimensional Euclidean geometry representations with many sampled points, cells, or mesh vertices, every particle is described by the same ordered spectral coefficient list. The ordering by \(n\) and \(m\) also gives the network direct access to a physically meaningful hierarchy of shape bandwidths, from global form to local roughness. The training particles are generated by random spherical-harmonic reconstruction under positivity and compactness constraints, following established morphology-generation procedures.\cite{Wei:2018simple,Liu:2023influence} Figure~\ref{fig:sh_representation}(a)--(c) summarizes this morphology encoding and its truncation convergence.

For the scattering spectrum, the electromagnetic transition matrix, or T-matrix, originally introduced by Waterman, is used.\cite{Waterman:1965matrix} At a fixed optical frequency, incident and scattered fields are expanded in spherical vector wave functions (SVWFs),
\begin{equation}
\Einc=\sum_{\alpha}c^{\mathrm{inc}}_{\alpha}\mathbf{W}^{\mathrm{reg}}_{\alpha},\qquad
\Esca=\sum_{\alpha}c^{\mathrm{sca}}_{\alpha}\mathbf{W}^{\mathrm{out}}_{\alpha},
\end{equation}
where the compound index \(\alpha=(n,m,P)\) labels angular momentum and polarization. The T-matrix defines the modal mapping
\begin{equation}
c^{\mathrm{sca}}_{\alpha}=\sum_{\beta}\Tmat_{\alpha\beta}c^{\mathrm{inc}}_{\beta}.
\label{eq:tmatrix}
\end{equation}
Equation~(\ref{eq:tmatrix}) defines the operator representation used in the framework. For a fixed particle geometry, material, background, and wavelength, \(\Tmat\) is a deterministic property of the scatterer itself; the incident direction, polarization, and wavefront determine only the incident modal vector \(c^{\mathrm{inc}}_{\beta}\) used to probe that property. Learning \(\Tmat\) is therefore different from learning one scattering image or one scalar cross section: the target retains phase, polarization mixing, and angular-momentum coupling information, and it can synthesize responses to arbitrary incident modal superpositions. Ground-truth T-matrices are computed with a boundary-element formulation using Rao--Wilton--Glisson basis functions on triangular surface meshes.\cite{Rao:1982electromagnetic,Jing:2023deep} Mesh tests show convergence for the 15th-order reconstructions when more than about 5000 triangular elements are used. Figure~\ref{fig:sh_representation}(d)--(f) illustrates this T-matrix ground-truth workflow, from surface meshing and RWG basis construction to mesh convergence and representative complex modal maps. In the present data set, all T-matrices are evaluated at 1064~nm, truncated at multipole order 10, and correspond to homogeneous particles in a uniform background. This fixed-frequency setting isolates the central morphology-to-operator relation before adding wavelength or material dispersion as additional input variables. The resulting target remains high dimensional because each output contains the real and imaginary parts of all retained input--output modal couplings. Predicting this object is more demanding than predicting a scalar efficiency, but it is also more informative because the same matrix represents the particle's intrinsic scattering response and can generate many downstream observables without recomputing the electromagnetic boundary-value problem.

The dual spectral structure suggests a compact learning architecture. Each particle is represented by standardized spherical-harmonic coefficients ordered by increasing degree. After removing or normalizing the lowest-order scale-dominated terms, the remaining coefficients are partitioned into non-overlapping groups and embedded as spectral tokens. A Transformer encoder processes these ordered tokens with positional encoding and multi-head self-attention, enabling the model to learn correlations among global and fine morphology modes. This is useful because scattering from an irregular particle is not determined by any single shape coefficient; low-order modes set the dominant multipoles, whereas higher-order perturbations break symmetry and introduce off-diagonal coupling between angular-momentum channels. Attention pooling produces a latent representation that is mapped by a multilayer perceptron to the vectorized complex T-matrix, expressed as concatenated real and imaginary parts,
\begin{equation}
\widehat{\Tmat}=F_{\Theta}\left(\{\tilde c_n^m\}_{n\le 15}\right),
\end{equation}
where \(F_{\Theta}\) denotes the trained spectral-token Transformer. The model is optimized on 50{,}000 irregular particles using standardized inputs and outputs, a variance-aware weighted mean-squared error, total-variation regularization on the reconstructed T-matrix, AdamW optimization, learning-rate scheduling, gradient clipping, and early stopping. The variance-aware loss prevents large, low-order T-matrix entries from completely dominating training while still preserving their physical weight in scattering observables. The regularization term suppresses unphysical high-frequency texture in the predicted modal maps, encouraging the output to remain smooth across neighboring modal indices without explicitly imposing a hard symmetry constraint. At inference, only the SH coefficient vector is required; no mesh generation or electromagnetic solve is performed. The predicted operator can then be post-processed by standard SVWF algebra to compute observables for any incident beam represented in the same modal basis.

\begin{figure*}[t]
    \centering
    \includegraphics[width=0.88\textwidth]{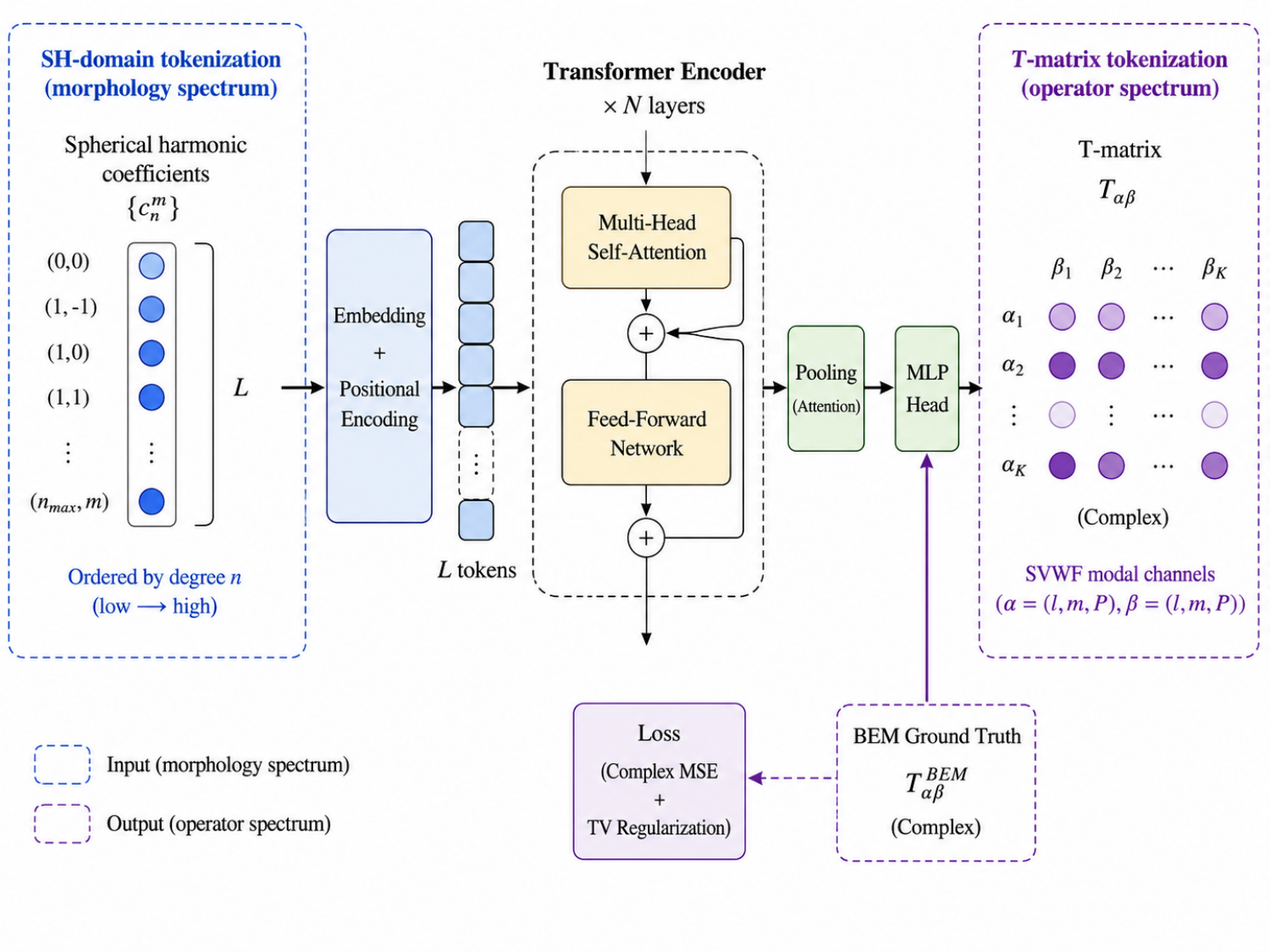}
    \caption{Spectral-token Transformer for morphology-to-operator learning. Ordered spherical-harmonic coefficients are grouped into spectral tokens, processed by self-attention layers, aggregated into a latent representation, and decoded into the real and imaginary parts of the finite-order complex T-matrix.}
    \label{fig:transformer}
\end{figure*}

The learned operator is first evaluated at the T-matrix level. This evaluation directly tests whether the model has captured the intrinsic response of the particle before any choice of incident beam or detector geometry is imposed. The global comparison between predicted and ground-truth elements yields a Pearson correlation coefficient of approximately 0.974 on the prediction set. More importantly, the modal maps show that the prediction preserves the block structure, diagonal dominance, and off-diagonal coupling texture of the ground-truth operator. These structures reflect angular-momentum coupling and symmetry breaking induced by shape irregularity, and they are not visible when only a single far-field pattern is learned. The agreement is therefore not merely an element-wise regression score; it indicates that the network has learned the organization of the scattering operator in the multipolar basis. The average error maps show larger errors in strongly coupled low-order blocks, consistent with the dominant role of low-order multipoles in Mie-regime observables. Errors in weak high-order entries are less visually prominent but remain important because they can influence fine angular interference, motivating the observable-level validation below.

\begin{figure*}[t]
    \centering
    \includegraphics[width=0.88\textwidth]{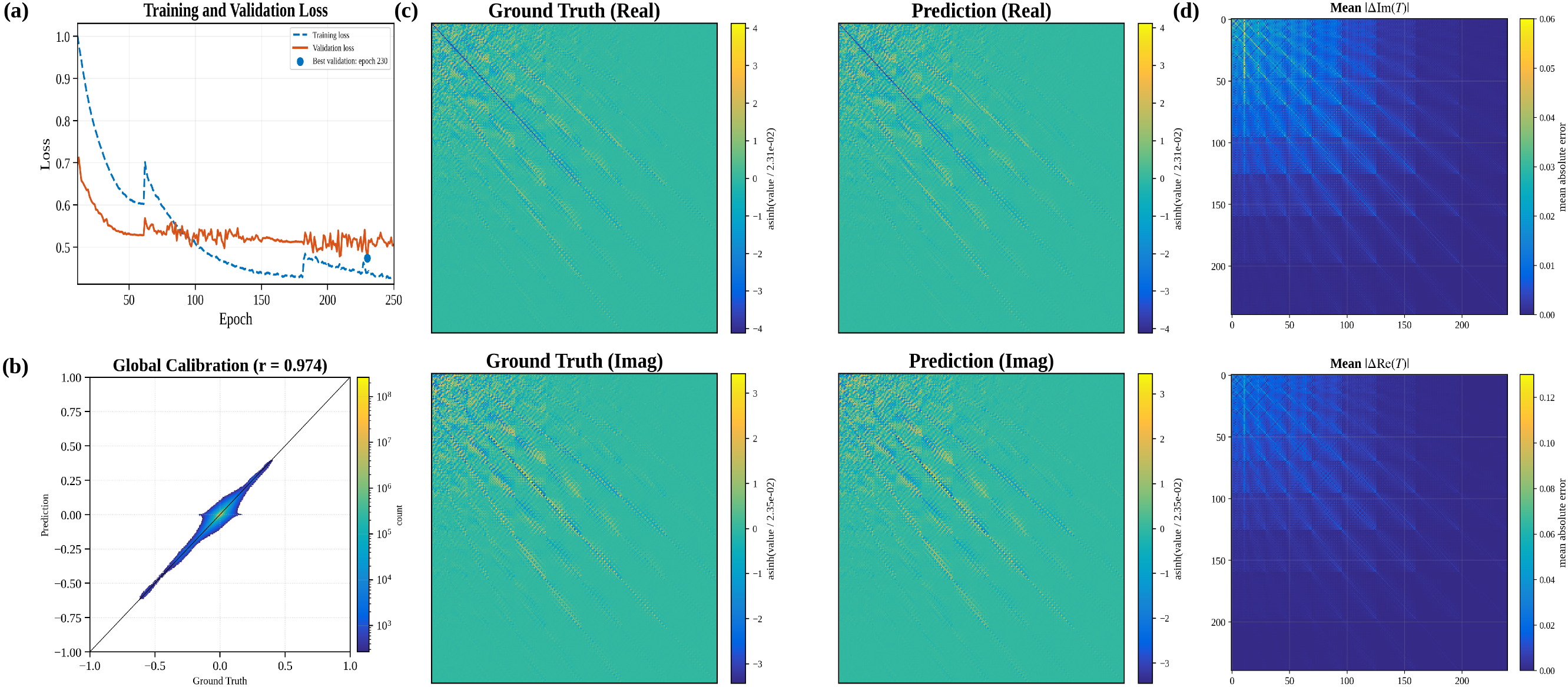}
    \caption{Learning the intrinsic T-matrix from spectral shape tokens. (a) Training and validation losses. (b) Element-wise agreement between ground-truth and predicted T-matrix entries. (c) Ground-truth and predicted real and imaginary modal maps for one validation particle. (d) Mean relative errors of real and imaginary T-matrix components over the prediction set.}
    \label{fig:tm_prediction}
\end{figure*}

\begin{figure*}[t]
    \centering
    \includegraphics[width=0.92\textwidth]{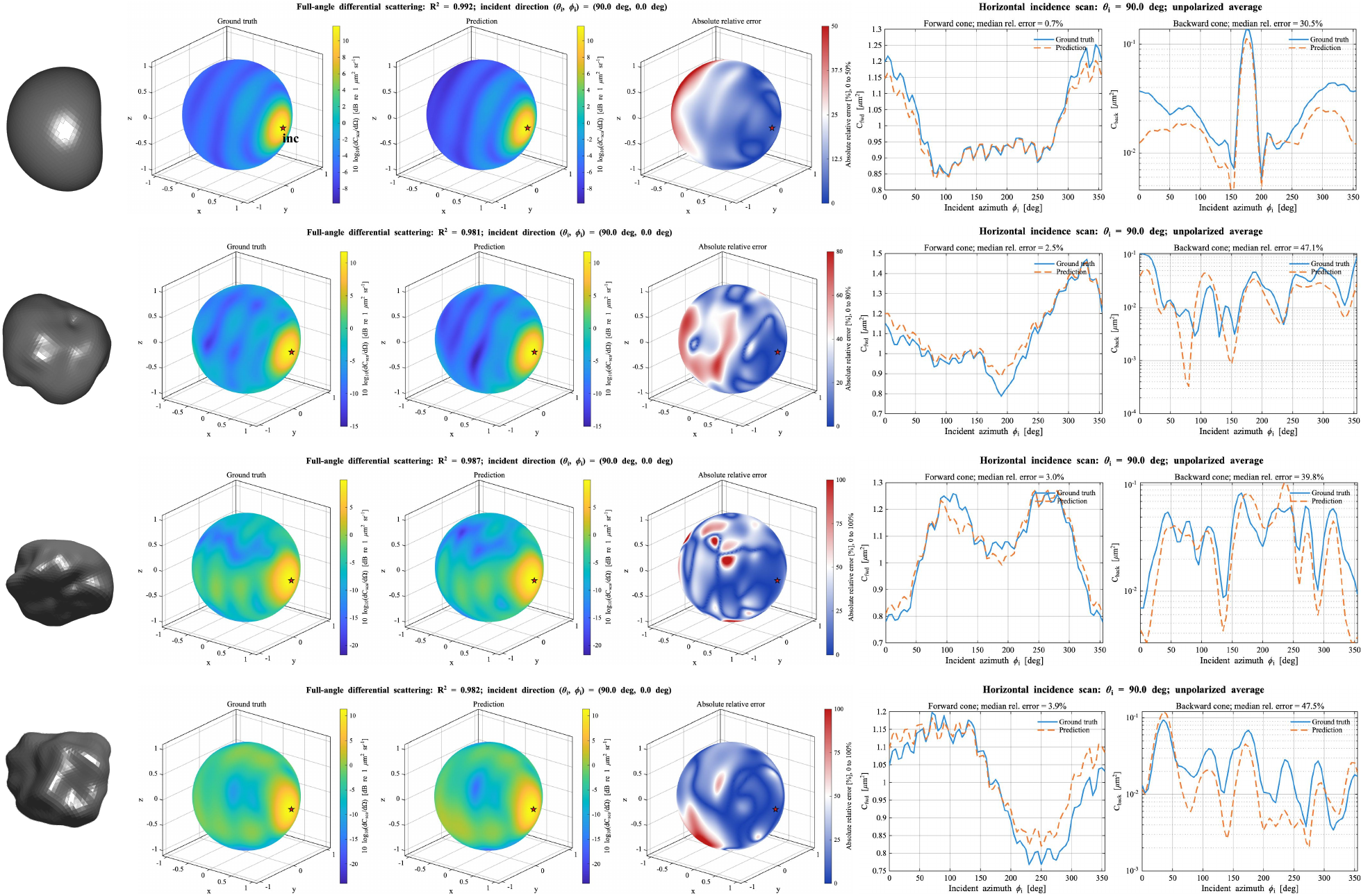}
    \caption{Physical scattering validation of predicted T-matrices at 1064~nm. Each row shows one irregular particle, full-angle differential scattering maps from the ground-truth and predicted T-matrices, the pointwise relative error, and forward/backward cone integrals during horizontal incidence scans. The same predicted T-matrix is reused throughout the scan.}
    \label{fig:scattering_validation}
\end{figure*}

Because an operator is only useful if it produces correct physical observables, predicted and reference T-matrices are next converted into differential scattering cross sections. Figure~\ref{fig:scattering_validation} compares four randomly selected irregular particles at 1064~nm. For fixed horizontal incidence, the predicted full-angle scattering maps reproduce the dominant lobes and interference features of the reference solutions, with displayed full-angle coefficients of determination of approximately 0.981--0.992. Local relative-error maps show that larger pointwise errors are confined to limited angular regions, typically where the reference intensity is small and the relative metric is correspondingly sensitive. The same predicted T-matrix is then reused while the incident azimuth is continuously scanned. In this test the incident direction enters only through the SVWF coefficients in Eq.~(\ref{eq:tmatrix}); the learned particle operator is unchanged. The predicted forward-cone integrals closely follow the ground-truth curves, showing that the dominant energy flow is recovered across illumination changes. The weaker backward-cone signals retain the main trends and peak positions despite larger relative sensitivity to phase and high-order modal details. This illumination reuse is the practical advantage of learning \(\Tmat\) rather than a fixed-output scattering image: one neural inference supplies a reusable optical object that can be interrogated repeatedly by inexpensive modal algebra. It also provides a more stringent test than a single-angle comparison, because errors that are hidden for one incidence can become visible when the incident coefficients rotate through the modal basis. The close agreement across the scan therefore indicates that the predicted operator has learned angular correlations, not only the response to one selected excitation.

Transfer beyond the random spherical-harmonic training distribution is finally assessed to determine whether the relation between spectral morphology and scattering operators generalizes. Generalization is evaluated on a 100-sample in-distribution baseline, four out-of-distribution synthetic shape families, and four real sand-particle materials from the Sand Atlas data set.\cite{Vego:2025sandatlas} For each sample, reference and predicted T-matrices are converted into full-angle differential scattering cross sections, and angular agreement is measured by \(R^2\). The in-distribution median is 0.983. Three synthetic out-of-distribution families remain highly predictable, with median \(R^2\) values of 0.920 for clipped spheroids, 0.884 for random convex polyhedra, and 0.940 for superquadrics. These shapes differ from the training generator but still have smooth or moderately band-limited radial structure, so their scattering can be represented within the learned bandwidth of the SH-to-T-matrix mapping. The radial triangular-mesh family is the clearest failure case, with median \(R^2=0.615\), because sharp protrusions and localized irregularities demand higher shape bandwidth and excite stronger high-order modal couplings. Real sand morphologies also transfer well: median \(R^2\) values are 0.934 for Caicos ooids, 0.851 for Hostun sand, 0.903 for LECA, and 0.890 for Ottawa sand. Their performance is significant because these particles are not generated by the training procedure and include non-ideal surface variations typical of natural materials. Together with the high prediction accuracy, this transfer supports the central claim that the model learns an intrinsic geometry-to-scattering relation in spectral space, rather than only memorizing a random shape generator or fitting a particular illumination condition. The weaker performance for radial triangular meshes identifies sharp, highly localized roughness as the main regime needing additional spectral resolution.

\begin{figure*}[t]
    \centering
    \includegraphics[width=0.9\textwidth]{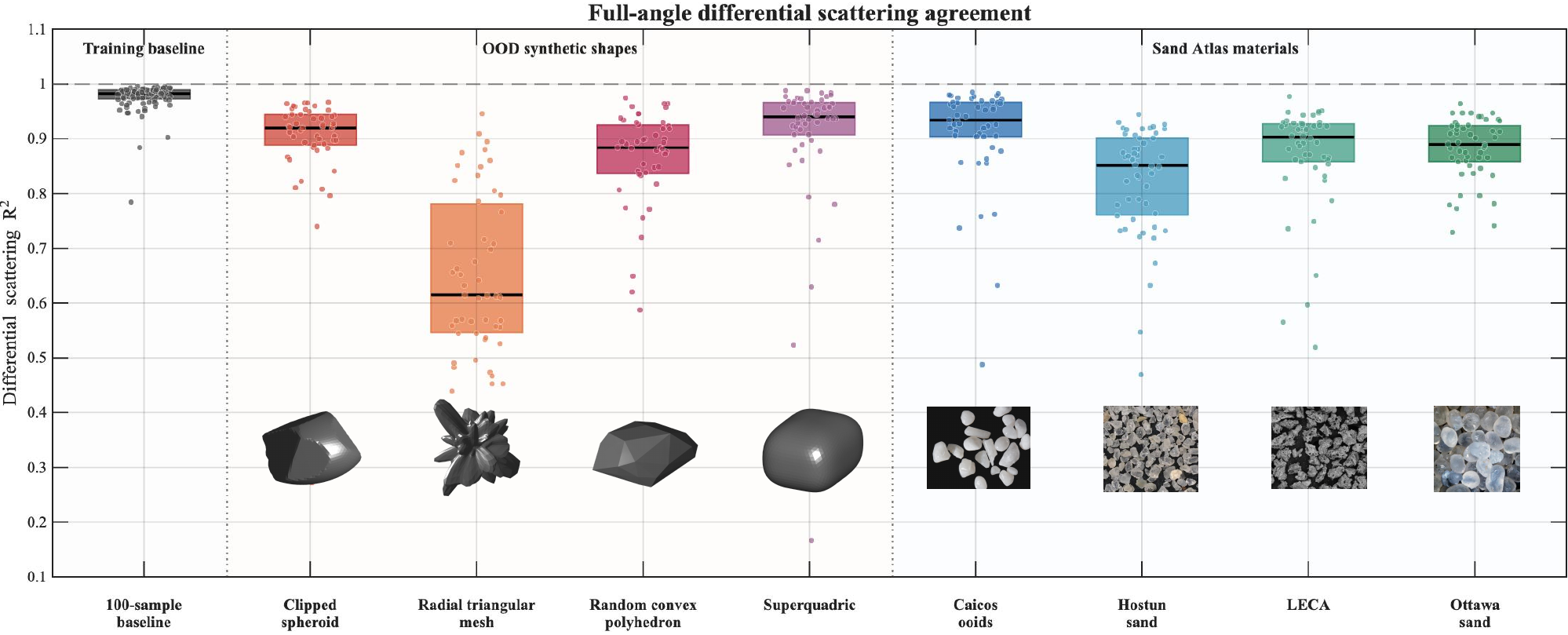}
    \caption{Generalization across in-distribution, out-of-distribution synthetic, and real sand-particle morphologies. Box plots show \(R^2\) between full-angle differential scattering cross sections computed from reference and predicted T-matrices at 1064~nm for fixed incidence \((\theta_i,\phi_i)=(90^\circ,0^\circ)\).}
    \label{fig:generalization}
\end{figure*}

The results demonstrate that the dual spectral-domain representation provides an effective way to learn optical scattering from complex particle geometry. Compressing each particle into 256 ordered spherical-harmonic coefficients removes the need for high-dimensional Euclidean geometry inputs such as voxel grids, point clouds, or meshes, while the complex T-matrix preserves the phase, polarization, and angular-momentum coupling information needed to reconstruct angle-resolved scattering. The learned operators reproduce both modal T-matrix structure and full-angle differential scattering observables, showing that the model is not only predicting a numerical array but recovering a physically useful scattering representation.

The generalization tests further clarify what has been learned. High accuracy for smooth or moderately band-limited out-of-distribution synthetic families and for natural sand-particle morphologies indicates that the spectral-token Transformer captures a transferable relation between geometry spectra and multipolar scattering responses. The weaker performance for radial triangular meshes is also informative: sharp protrusions and highly localized roughness exceed the effective shape bandwidth of the present spherical-harmonic representation and excite scattering channels that require additional spectral resolution. Thus, the main limitation is not the operator-learning concept itself, but the bandwidth and geometric coverage of the morphology descriptor used here.

Several extensions follow naturally from this conclusion. Multiple spherical charts or hybrid spectral--mesh descriptors could extend the approach beyond star-shaped particles and improve treatment of localized roughness. Broadband material tokens could incorporate wavelength and dispersion, while reciprocity, passivity, and energy-conservation constraints could make the predicted T-matrices more robust outside the training distribution. Uncertainty estimation would also be valuable because one predicted T-matrix can be reused for many incident fields and observation geometries. Overall, the present results establish spectral-domain operator learning as a compact and transferable route for efficient, reusable optical scattering prediction of complex three-dimensional particles.

\begin{acknowledgments}
This research is financially supported by the National Natural Science Foundation of China (Grant Nos. U24A2087 and 52574155), the Guangdong Basic and Applied Basic Research Foundation (No. 2024A1515011952), and the Stable Support Program Project of Shenzhen Municipal Science and Technology Innovation Committee (No. 202311201748\allowbreak39003).
\end{acknowledgments}

\noindent\textbf{Conflict of Interest.} The authors have no conflicts to disclose.

\noindent\textbf{Author Contributions.} Daize Li: Conceptualization, methodology, software, validation, visualization, writing--original draft. Jiafu Shen: Computation, validation, visualization. Yifei Liu: Conceptualization, supervision, funding acquisition, writing--review and editing. Bonan Zhang: Methodology, writing--review and editing. Heping Xie: Supervision, funding acquisition.

\noindent\textbf{Data Availability.} The data that support the findings of this study are available from the corresponding author upon reasonable request.

\bibliography{ref}

\end{document}